\documentclass[prd,
,superscriptaddress,
nofootinbib,%
tightenlines ]{revtex4}
\usepackage{epsfig}
\usepackage[colorlinks=true,linkcolor=blue,urlcolor=blue,citecolor=blue]{hyperref}
\usepackage{amsfonts}
\usepackage{amssymb}
\usepackage{amsmath}
\usepackage{slashed}
\usepackage{color}

\newcommand{\ben}{\begin{displaymath}}
\newcommand{\een}{\end{displaymath}}
\newcommand{\be}{\begin{equation}}
\newcommand{\ee}{\end{equation}}
\newcommand{\bea}{\begin{eqnarray}}
\newcommand{\eea}{\end{eqnarray}}

\begin{document}
\title{Gravitational form factors of the $Z$-boson}
\author{P.~Bei\ss ner}
 \affiliation{Institut f\"ur Theoretische Physik II, Ruhr-Universit\"at Bochum,  D-44780 Bochum,
 Germany}
 \author{J.~Yu.~Panteleeva}
\address{Institut f\"ur Theoretische Physik II, Ruhr-Universit\"at Bochum,  D-44780 Bochum, Germany}
 \author{B.-D.~Sun}
 \affiliation{Institut f\"ur Theoretische Physik II, Ruhr-Universit\"at Bochum,  D-44780 Bochum,
 Germany}
\author{E.~Epelbaum}
 \affiliation{Institut f\"ur Theoretische Physik II, Ruhr-Universit\"at Bochum,  D-44780 Bochum,
 Germany}
\author{J.~Gegelia}
 \affiliation{Institut f\"ur Theoretische Physik II, Ruhr-Universit\"at Bochum,  D-44780 Bochum,
 Germany}
 \affiliation{Tbilisi State
University, 0186 Tbilisi, Georgia}

\date{30 January, 2026}

\begin{abstract}

Matrix elements of the energy-momentum tensor for one-particle states of the $Z$-boson are parameterized in terms of gravitational form factors.
One-loop order electroweak corrections to these quantities are calculated.
Renormalization and physical interpretation of the obtained results are discussed. 

\end{abstract}

\maketitle

\section{Introduction}

Our everyday perception of the surrounding world is dominated by the electromagnetic interaction. Three-dimensional images of macroscopic objects so familiar to us  are generated by photons scattered on them. This macroscopic picture, extrapolated to quantum mechanical systems like atoms, leads to our approximate classical interpretation of objects which are by no means classical. Being well aware of this fact, it is still comforting to have an intuitively clear picture of, e.g., the charge distributions of composite particles, like hadrons. 
There exist, however, electromagnetically truly neutral particles that cannot be probed by photons. 
Another long-range force, gravitation, can be used to probe the structure of such systems.  
It does not seem feasible to measure scattering processes of particles off external
gravitational sources, however theoretical investigations can be carried out, e.g., in the framework of the standard model. 
	In quantum field theoretical formalism, gravitational scattering processes in the one-graviton-exchange approximation are
described by diagrams in which a graviton couples 
to one-particle matrix elements of the energy-momentum tensor (EMT).   
Such matrix elements and the corresponding gravitational form
factors (GFF) \cite{Kobzarev:1962wt,Pagels:1966zza} have been
extensively studied for various systems in recent years.
In particular, the GFFs of hadrons have attracted much attention, see,
e.g.,
Refs.~\cite{Polyakov:1999gs,Polyakov:2002yz,Polyakov:2018zvc,Ji:1996ek,Hudson:2017xug,Nature,Lorce:2025oot,Kumano:2017lhr,Kumericki:2019ddg,Shanahan:2018nnv,Shanahan:2018pib,Diehl:2006ya,Avelino:2019esh,Belitsky:2002jp,Lorce:2017wkb,Schweitzer:2019kkd,Goeke:2007fp,Polyakov:2020rzq,Alharazin:2020yjv,Gegelia:2021wnj,Epelbaum:2021ahi,Alharazin:2022wjj,Alharazin:2023zzc,Alharazin:2023uhr}. Calculations
of the gravitational structure due to strong interaction effects have been
performed using different methods. The increased interest to this
topic is driven by the fact  that GFFs of hadrons can be (indirectly)  extracted from
experiment. 
Compared to the gravitational force,  electromagnetic and weak
interactions are also very strong and, therefore, it makes sense to
consider the electroweak corrections to matrix elements of the EMT operator.
Corrections due to the electromagnetic interaction lead to infrared
divergences \cite{Kubis:1999db,BjerrumBohr:2002kt,Varma:2020crx,Freese:2022jlu} that require taking into account emission of soft photons. However, for neutral particles, this problem does not occur.  
One-loop electroweak corrections to the
matrix elements of the EMT operator for the Higgs boson have been calculated in Ref.~\cite{Beissner:2025nmg}. 
In this work, we calculate the one-loop corrections to the GFFs of the $Z$-boson. 
Following the modern point of view, we consider the standard model as a leading order approximation to an effective field theory (EFT) \cite{Weinberg:1996kw}. Gravitation can be incorporated in this framework by constructing most general effective Lagrangian invariant under general coordinate transformations \cite{Donoghue:1994dn}. 
All ultraviolet (UV) divergences stemming from loop contributions to the matrix elements of the EMT operator are cancelled by counter terms generated by the Lagrangian of this EFT. 

In the next section, we parameterize matrix elements of the EMT operator for $Z$-boson one-particle states in terms of GFFs and specify the details of calculation of one-loop corrections to these quantities.  
The results of our work are
summarized in  section~\ref{conclusions}. The appendices contain the definition of the required one-loop integrals and expressions of the calculated GFFs at vanishing transferred momentum squared.

\section{One-loop corrections to the gravitational form factors of the  $Z$-boson}
\label{EMFFs}

	To calculate the one-particle-state matrix elements of the EMT operator for the $Z$-boson we use the Lagrangian and the Feynman rules of the electroweak theory as specified in Ref.~\cite{Aoki:1982ed}.
To obtain the EMT operator we consider the Lagrangian of the electroweak theory coupled to gravitational field, use the formulas of Ref.~\cite{Birrell:1982ix} for relating the EMT operator to the Lagrangian
and exploit the results of Ref.~\cite{Freedman:1974gs} for a non-Abelian gauge theory with spontaneous symmetry breaking. 

\begin{figure}[t]
\begin{center}
\epsfig{file=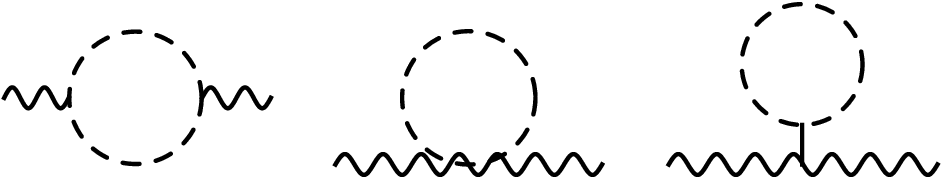,scale=0.5}
\caption{Topologies of one-loop diagrams contributing to the self-energy of the $Z$-boson. Wiggly lines correspond to $Z$-bosons while dashed lines represent vector bosons, fermions, Higgs bosons, Goldstone bosons and Faddeev-Popov ghosts.}
\label{Ziggs_SE}
\end{center}
\vspace{-5mm}
\end{figure}

\medskip

We parameterize the matrix elements of the EMT operator for one-particle states of spin-1 particles in terms of GFFs as follows:
\begin{eqnarray}
\langle p', \sigma'| T_{\mu\nu}| p,\sigma \rangle &=& \epsilon^{* \alpha'} (p',\sigma') \epsilon^\alpha (p,\sigma) 
\Biggl\{ 2 P_\mu P_\nu  \left( -\eta_{\alpha\alpha'} A_0(t)  + \frac{P_\alpha P_{\alpha'}} {M_V^2}\,A_1(t)   \right) \nonumber\\
&& + 2 \left[ P_\mu ( \eta_{\nu \alpha'} P_{\alpha}+\eta_{\nu\alpha} P_{\alpha'}) + P_\nu (\eta_{\mu \alpha'} P_{\alpha}+\eta_{\mu\alpha} P_{\alpha'} )\right] J(t)  \nonumber\\
&+&  \frac{1}{2} (q_\mu q_\nu-\eta_{\mu\nu} q^2) \left( \eta_{\alpha\alpha'}D_0(t) + \frac{P_\alpha P_{\alpha'}} {M_V^2}\, D_1(t) \right) \nonumber\\
&+& \biggl[ \frac{1}{2} \, (\eta_{\mu \alpha}  \eta_{\nu \alpha'} + \eta_{\mu \alpha'}  \eta_{\nu \alpha}  )q^2 
- ( \eta_{\nu \alpha'} q_\mu   + \eta_{\mu \alpha'}  q_\nu  ) P_\alpha  \nonumber\\
&+& ( \eta_{\nu \alpha} q_\mu   + \eta_{\mu \alpha}  q_\nu  ) P_{\alpha'} -4 \, \eta_{\mu\nu} P_\alpha P_{\alpha'}
\biggr] E(t) \nonumber\\ 
&+& \frac{ i } {M_V^2}\, \, \epsilon_{\alpha \alpha'\lambda\sigma } P^\lambda q^\sigma  \left[ P_\mu P_\nu \, F_1(t) +  (q_\mu q_\nu-\eta_{\mu\nu} q^2) \, F_2(t) \right]  + 
i \,  \left(P_\nu \epsilon_{\alpha \alpha' \mu \sigma}  q^\sigma  + P_\mu \epsilon_{\alpha \alpha' \nu\sigma } q^\sigma \right) \, F_3(t) \nonumber\\
&+& 
 \frac{i } {M_V^2}\,  \left(\epsilon_{\alpha \mu\lambda\sigma} P\lambda q^\sigma  P_\nu P_{\alpha'} +\epsilon_{\alpha \nu\lambda\sigma} P^\lambda q^\sigma  P_\mu P_{\alpha'}-
\epsilon_{\alpha' \mu\lambda\sigma} P^\lambda q^\sigma  P_\nu P_{\alpha} - \epsilon_{\alpha' \nu\lambda\sigma} P^\lambda q^\sigma  P_\mu P_{\alpha} \right) \, F_4(t) 
\nonumber\\
&+& 
 \frac{i } {M_V^2}\, \left[ \epsilon_{\alpha \mu\lambda\sigma} P^\lambda q^\sigma  \left(q_\nu q_{\alpha'} -q^2 \eta_{\nu \alpha'}\right)+\epsilon_{\alpha \nu\lambda\sigma} P^\lambda q^\sigma  \left(q_\mu q_{\alpha'} -q^2 \eta_{\mu \alpha'}\right) \right. \nonumber\\
&-& \left. 
\epsilon_{\alpha' \mu\lambda\sigma} P^\lambda q^\sigma  \left(q_\nu q_{\alpha} -q^2 \eta_{\nu \alpha}\right) 
- \epsilon_{\alpha' \nu\lambda\sigma} P^\lambda q^\sigma  \left(q_\mu q_{\alpha} -q^2 \eta_{\mu \alpha}\right)
 \right] \, F_5(t)
\Biggr\} \,,
\label{RGFFsdef}
\end{eqnarray}
where $t=q^2=(p' - p)^2$, $P=(p' + p)/2$, and $M_V$ is a mass parameter introduced to make the form factors dimensionless (it is convenient to choose $M_V=M$, where $M$ is the mass of the considered spin-1 system). The polarization
vector $\epsilon_\alpha(p,\sigma)$ satisfies the condition
\begin{equation}
\sum_\sigma \epsilon_\alpha(p,\sigma) \epsilon_\beta(p,\sigma) =-\eta_{\alpha\beta}+\frac{p^\alpha p^\beta}{M^2}\,.
\label{sumofpols}
\end{equation} 
Notice that this parameterization differs from those of Refs.~\cite{Holstein:2006ud,Polyakov:2019lbq,Cotogno:2019vjb} which consider systems with even parity. 

\medskip

As the $Z$-boson is an unstable particle, we extract its gravitational form factors from the
residue at the complex double-pole of the three-point correlation function of  the EMT operator 
and two $Z$-boson fields \cite{Gegelia:2010nmt}.
To apply the LSZ formalism to the correlation function we need to calculate the one-loop
contributions to the pole position  --  $\delta z_1$, and to the residue of the dressed propagator of the $Z$-boson -- $\delta  Z_1$.
Topologies of one-loop diagrams contributing to the self-energy of the $Z$-boson are shown in Fig.~\ref{Ziggs_SE}. 
The corresponding expressions for $\delta z_1$ and $\delta  Z_1$ are given in appendix \ref{AppB}. 

\medskip

In the calculation of the one-loop diagrams we apply dimensional
regularization (see, e.g., Ref.~\cite{Collins:1984xc}) with $D$
spacetime dimensions and use the program
FeynCalc
\cite{Mertig:1990an,Shtabovenko:2016sxi,Shtabovenko:2023idz}. The
results of our calculations are expressed in terms of scalar integrals
defined in appendix  \ref{AppA}. 

\medskip

\begin{figure}[t]
\begin{center}
\epsfig{file=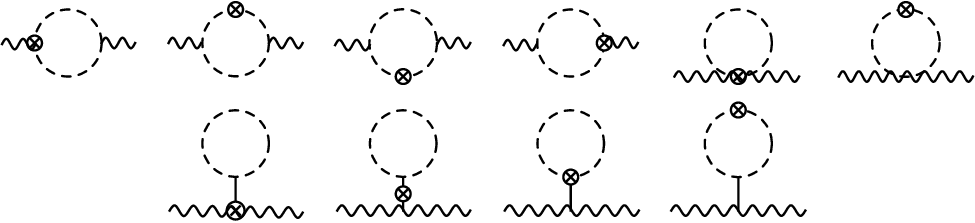,scale=0.8}
\caption{Topologies of one-loop diagrams contributing to the
  three-point vertex function of  the EMT operator and two $Z$-boson
  fields. The crosses stand for the EMT insertions and the wiggly lines refer
  to $Z$-bosons. The dashed lines represent vector bosons, fermions, Higgs bosons, Goldstone bosons and Faddeev-Popov ghosts.}
\label{Ziggs_EMT}
\end{center}
\vspace{-5mm}
\end{figure}

The one-loop topologies contributing to the amputated three-point function are
shown in Fig.~\ref{Ziggs_EMT}. 
By calculating these diagrams and using the LSZ formalism we extract the GFFs, whose
expressions are too lengthy to be given explicitly.\footnote{The corresponding {\it Mathematica} notebook is available upon request.} 
Our calculated results satisfy the condition $A_0(0)=1$, $J(0)=1$ as expected from general considerations. 
The expressions for other GFFs at $q^2=0$ are specified in appendix \ref{GFFs0}.

\begin{figure}[t]
\begin{center}
\epsfig{file=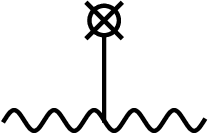,scale=0.6}
\caption{Counter term diagram cancelling the divergence of the one-loop contribution to the matrix element of the EMT operator.
The vertex with cross stands for the counter term, and the wiggly and straight solid lines refer to $Z$- and Higgs bosons, respectively.}
\label{CTD}
\end{center}
\vspace{-5mm}
\end{figure}

All one-loop contributions to GFFs are finite except $D_0$, whose divergent part has the form:
\begin{equation}
   D_0^{div}(t)  =  -\frac{e^2 M_Z^4 \left(-3 M_H^2-2 m_n^2+6 M_W^2+3 M_Z^2\right)}{24 \pi ^2
   (D-4) M_W^2 \left(M_H^2-t\right) \left(M_W^2-M_Z^2\right)}
    \,,
\label{DivParts}
\end{equation} 
where $M_Z$, $M_W$ and $M_H$ are the masses of the $Z$-, $W$- and Higgs bosons, respectively.
The UV divergence in Eq.~(\ref{DivParts}) is cancelled by the counter term diagram of Fig.~\ref{CTD} with a counter term  generated by an interaction term of the effective Lagrangian
\begin{equation}
c \, R\, \Phi^\dagger \Phi \,,
\label{ctTerm2}
\end{equation}
where $c$ is a coupling constant and $R$ is the scalar curvature. We use the notations of Ref.~\cite{Aoki:1982ed}, where the Higgs field is expressed in terms of hermitian fields $\phi$, $\chi_1$, $\chi_2$, $\chi_3$ as
\begin{equation}
\Phi = \frac{1}{\sqrt{2}}\left(
\begin{array}{c}
 i \chi_1+\chi_2    \\
v + \phi- i \chi_3
\end{array}
\right)\,,
\label{PhiDef}
\end{equation}
with the constant $v$ corresponding to the minimal point of the potential $V(\Phi)$.

The contribution to the EMT stemming from the term in Eq.~(\ref{ctTerm2}) in flat spacetime has the form:
\begin{equation}
T_{\mu\nu}= 2 \, c \left( \eta_{\mu\nu} \partial^2-\partial_\mu\partial_\nu \right)\left( \Phi^\dagger \Phi \right).
\label{EMTc1c2}
\end{equation}
By substituting Eq.~(\ref{PhiDef}) in Eq.~(\ref{EMTc1c2}) and splitting the bare coupling into the renormalized one and the counter terms $c=c_R +\hbar \delta c_1 +\hbar^2 \delta c_2 + ... $
we generate the vertex of external graviton interacting with a single scalar which contributes to Fig.~\ref{CTD}. By adjusting the counter term $\delta c_1$ we cancel the divergence given in Eq.~(\ref{DivParts}). We have checked explicitly that the same value of $\delta c_1$ cancels the analogous divergence for the Higgs boson, the one-loop corrections to GFFs of which have been calculated in Ref.~\cite{Beissner:2025nmg}.

\medskip

\medskip

By considering matrix elements of the EMT operator in localized one-particle states, the GFFs can be related to corresponding spatial distributions.  
In the case of the $Z$-boson, the static
approximation is not applicable as the characteristic gravitational radii of the system (given by GFFs and their derivatives at vanishing momentum transfer) are smaller than its Compton wavelength \cite{Jaffe:2020ebz}.
Therefore, we need to consider sharply localized states with the
sizes of the wave packets chosen much smaller than the Compton wavelength \cite{Panteleeva:2023evj,Epelbaum:2022fjc}. 
Such wave packet states are dominated by high momenta, and the spatial distribution corresponding to the following linear combination of GFFs: 
\begin{equation}
 \mathcal{E}_0(q^2)=A_0(q^2)+\dfrac{q^2}{12 m^2}A_1(q^2)
-\dfrac{q^2}{12m^2}\left(4J(q^2)-2E( q^2)-2A_0( q^2) - A_1(q^2)\dfrac{q^2}{4 m^2}\right)
\label{defineE0}
\end{equation}
can be interpreted as the energy distribution \cite{Panteleeva:2023evj}.
The corresponding mean square energy radius is given by  
\begin{equation}
r_0^2= 4\, \frac{ d {\cal E}_0
}{dq^2} \big|_{q^2 = 0}\,. 
\label{defr2}
\end{equation} 
Our calculations lead to the following result for this quantity:
\begin{eqnarray}
r^2_{0} & = & \frac{ e^2}{432 \pi ^2 M_W^2 M_Z^6
   \left(M_W^2-M_Z^2\right) \left(M_Z^2-4 m_n^2\right){}^2}  \left(12 \left(5 M_Z^6+4 M_V^2 M_Z^4+8 \left(2 M_V^2+M_Z^2\right)
   \left(M_Z^2-M_W^2\right) Q_n s_n M_Z^2 \right.\right. \nonumber\\
   &+& \left.\left. 16 \left(M_W^2-M_Z^2\right){}^2 \left(2
   M_V^2+M_Z^2\right) Q_n^2\right) m_n^6-4 M_Z^2 \left(5 M_V^2-14 M_Z^2\right)
   \left(M_Z^4+4 \left(M_Z^2-M_W^2\right) Q_n s_n M_Z^2 \right.\right. \nonumber\\
   &+& \left.\left. 8
   \left(M_W^2-M_Z^2\right){}^2 Q_n^2\right) m_n^4-2 M_Z^4 \left(M_V^2+32
   M_Z^2\right) \left(M_Z^4+4 \left(M_Z^2-M_W^2\right) Q_n s_n M_Z^2+8
   \left(M_W^2-M_Z^2\right){}^2 Q_n^2\right) m_n^2 \right. \nonumber\\
   &+& \left. M_Z^6 \left(M_V^2+11
   M_Z^2\right) \left(M_Z^4+4 \left(M_Z^2-M_W^2\right) Q_n s_n M_Z^2+8
   \left(M_W^2-M_Z^2\right){}^2 Q_n^2\right)\right) \nonumber\\
   &+& \frac{ A_0\left(m_n^2\right) e^2}{72 \pi ^2
   M_W^2 M_Z^6 \left(M_W^2-M_Z^2\right) \left(M_Z^2-4
   m_n^2\right){}^2} 
    \left(2
   \left(M_Z^4+4 \left(M_Z^2-M_W^2\right) Q_n s_n M_Z^2+8
   \left(M_W^2-M_Z^2\right){}^2 Q_n^2\right) M_Z^6 \right.\nonumber\\
   &+& \left. m_n^2 \left(-7 M_Z^8+2 M_V^2
   M_Z^6-8 \left(M_V^2-5 M_Z^2\right) \left(M_W^2-M_Z^2\right) Q_n s_n M_Z^4+16
   \left(M_V^2-5 M_Z^2\right) \left(M_W^2-M_Z^2\right){}^2 Q_n^2 M_Z^2\right) \right. \nonumber\\
   &-& \left. 2
   m_n^4 \left(5 M_Z^6+4 M_V^2 M_Z^4+8 \left(2 M_V^2+M_Z^2\right)
   \left(M_Z^2-M_W^2\right) Q_n s_n M_Z^2+16 \left(M_W^2-M_Z^2\right){}^2 \left(2
   M_V^2+M_Z^2\right) Q_n^2\right)\right) \nonumber\\
   &-& \frac{B_0\left(M_Z^2,m_n^2,m_n^2\right) e^2}{72 \pi ^2
   M_W^2 M_Z^6 \left(M_W^2-M_Z^2\right) \left(M_Z^2-4 m_n^2\right){}^2} \left(-2 \left(5 M_Z^6+4 M_V^2 M_Z^4+8 \left(2
   M_V^2+M_Z^2\right) \left(M_Z^2-M_W^2\right) Q_n s_n M_Z^2 \right. \right. \nonumber\\
   &+& \left.\left. 16
   \left(M_W^2-M_Z^2\right){}^2 \left(2 M_V^2+M_Z^2\right) Q_n^2\right)
   m_n^6-\left(7 M_Z^8-2 M_V^2 M_Z^6+8 \left(M_V^2-5 M_Z^2\right)
   \left(M_W^2-M_Z^2\right) Q_n s_n M_Z^4  \right. \right. \nonumber\\
   &-& \left. \left. 16 \left(M_V^2-5 M_Z^2\right)
   \left(M_W^2-M_Z^2\right){}^2 Q_n^2 M_Z^2\right) m_n^4+2 M_Z^6 \left(M_Z^4+4
   \left(M_Z^2-M_W^2\right) Q_n s_n M_Z^2+8 \left(M_W^2-M_Z^2\right){}^2
   Q_n^2\right) m_n^2\right) 
\nonumber\\
   &-& \frac{e^2}{864 \pi ^2 M_H^2 M_W^2
   M_Z^6 \left(M_H^2-4 M_Z^2\right){}^2 \left(4 M_W^4-5 M_Z^2
   M_W^2+M_Z^4\right)}  \left(144 \left(2 M_V^2+M_Z^2\right) \left(M_H^3-4 M_H M_Z^2\right){}^2
   M_W^8. \right. \nonumber\\
   &-& \left. 72 M_H^2 M_Z^2 \left(2 M_V^2-17 M_Z^2\right) \left(M_H^2-4
   M_Z^2\right){}^2 M_W^6+16 M_Z^4 \left(M_V^2+26 M_Z^2\right) \left(M_H^3-4 M_H
   M_Z^2\right){}^2 M_W^4 \right.\nonumber\\
   &+& \left. 6 M_Z^4 \left(96 M_Z^{10}-368 M_H^2 M_Z^8+8 \left(19
   M_H^4+4 M_V^2 M_H^2\right) M_Z^6+\left(32 M_H^4 M_V^2-7 M_H^6\right) M_Z^4-2
   \left(M_H^8+13 M_V^2 M_H^6\right) M_Z^2  \right. \right.\nonumber\\
   &+& \left.\left.  4 M_H^8 M_V^2\right) M_W^2-M_Z^6
   \left(144 M_Z^{10}-208 M_H^2 M_Z^8+8 \left(7 M_H^4+8 M_V^2 M_H^2\right)
   M_Z^6+\left(11 M_H^6+40 M_V^2 M_H^4\right) M_Z^4 \right. \right.\nonumber\\
   &-& \left. \left. \left(3 M_H^8+38 M_V^2
   M_H^6\right) M_Z^2+6 M_H^8 M_V^2\right)\right) \nonumber\\
   &+& \frac{  A_0\left(M_H^2\right) e^2}{144 \pi ^2 M_H^2 M_W^2 M_Z^4 \left(M_H^2-4
   M_Z^2\right){}^2 \left(M_W^2-M_Z^2\right)}  \left(84 M_Z^{10}-60 M_H^2 M_Z^8+5 \left(5 M_H^4+8
   M_V^2 M_H^2\right) M_Z^6 \right. \nonumber\\
   &-& \left. 2 \left(4 M_H^6+25 M_V^2 M_H^4\right)
   M_Z^4+\left(M_H^8+18 M_V^2 M_H^6\right) M_Z^2-2 M_H^8 M_V^2\right)
  \nonumber\\
   &+& \frac{A_0\left(M_W^2\right) e^2}{72 \pi ^2 M_W^2 M_Z^6
   \left(4 M_W^4-5 M_Z^2 M_W^2+M_Z^4\right)} \left(5 M_Z^8-79 M_W^2
   M_Z^6+104 M_W^4 M_Z^4+12 M_W^6 M_Z^2 \right. \nonumber\\
   &+& \left. 2 M_V^2 M_W^2 \left(12 M_W^4-4 M_Z^2
   M_W^2+M_Z^4\right)\right) \nonumber\\
   &-& \frac{A_0\left(M_Z^2\right) e^2}{144 \pi ^2 M_H^2 M_W^2 M_Z^4
   \left(M_H^2-4 M_Z^2\right){}^2 \left(M_W^2-M_Z^2\right)} \left(24 M_Z^{10}-20 M_H^2
   M_Z^8+2 \left(7 M_H^4+8 M_V^2 M_H^2\right) M_Z^6 \right. \nonumber\\
   &-& \left. \left(7 M_H^6+36 M_V^2
   M_H^4\right) M_Z^4+\left(M_H^8+16 M_V^2 M_H^6\right) M_Z^2-2 M_H^8
   M_V^2\right)  \nonumber\\
   &-& \frac{\left(5
   M_Z^8-79 M_W^2 M_Z^6+104 M_W^4 M_Z^4+12 M_W^6 M_Z^2+2 M_V^2 M_W^2 \left(12
   M_W^4-4 M_Z^2 M_W^2+M_Z^4\right)\right) B_0\left(M_Z^2,M_W^2,M_W^2\right)
   e^2}{72 \pi ^2 M_Z^6 \left(4 M_W^4-5 M_Z^2 M_W^2+M_Z^4\right)} \nonumber\\
   &+& \frac{B_0\left(M_Z^2,M_Z^2,M_H^2\right) e^2}{144 \pi ^2 M_H^2 M_W^2 M_Z^4
   \left(M_H^2-4 M_Z^2\right){}^2 \left(M_W^2-M_Z^2\right)}  \left(24
   M_Z^{12}-104 M_H^2 M_Z^{10}+2 \left(37 M_H^4+8 M_V^2 M_H^2\right) M_Z^8 \right. \nonumber\\
   &-& \left. 4
   \left(8 M_H^6+19 M_V^2 M_H^4\right) M_Z^6+\left(9 M_H^8+66 M_V^2 M_H^6\right)
   M_Z^4-M_H^8 \left(M_H^2+20 M_V^2\right) M_Z^2+2 M_H^{10} M_V^2\right)
    \,.
\label{Rs}
\end{eqnarray} 
In Eq.~(\ref{Rs}) as well as in the expressions of the GFFs for vanishing $t$ specified in appendix  \ref{GFFs0},  the fermionic contributions need to be summed up over the index $n$ which enumerates all quarks and leptons, and the corresponding signs $s_n$ and charges $Q_n$ need to be fixed according to the Feynman rules of Ref.~\cite{Aoki:1982ed} ($s_n=-1$ for upper- and $s_n=1$ for lower-fermions in doublets of quarks and leptons).

Ultraviolet divergences of loop functions contained in Eq.~(\ref{Rs}) cancel each other exactly so that the mean-square radius is finite.  Substituting numerical values of various parameters from Ref.~\cite{ParticleDataGroup:2024cfk} we estimate:
\begin{equation}
r^2 
= [2.36\times 10^{-8} + (-1.41 \times10^{-8}+8.2\times10^{-12} i)] \ {\rm fm}^2\,,
\label{MSQERNUM}
\end{equation}
where the number in the round brackets refers to the contribution of
fermions.  The expression of Eq.~(\ref{MSQERNUM}) is a complex number, as usual for unstable systems (see, e.g., Refs.~\cite{Gegelia:2010nmt,Hacker:2006gu,Pascalutsa:2004je,Beissner:2025nmg}).

\section{Summary}
\label{conclusions}

Due to our lack of intuition for the microscopic world we often describe microscopic quantum systems using notions and terminology of classical physics.  
For example, we observe that in first approximation, the deuteron interacts with an electron as if it was a classical charge distributed in space with some mean square radius. In this sense, the electromagnetic form factors of quantum systems are related to spatial densities characterizing the internal structure of these systems. 
For electromagnetically truly neutral particles, the gravitational interaction can be used to play analogous role. Internal structure of systems as probed by gravitational interaction
is encoded in one-particle matrix elements of the EMT operator, parameterized in terms of the GFFs. Compared to the gravitational force, the electroweak interaction is very strong.
Therefore, it makes sense to calculate the electroweak corrections to GFFs of neutral particles.    
In this work we have calculated the one-loop electroweak corrections to the GFFs of the $Z$-boson. We found that matrix elements of the EMT operator for one-particle states of the $Z$-boson contain ultraviolet divergences, which are cancelled by a counter term generated by the most general Lagrangian of the EFT of the standard model coupled to gravitational field, invariant under general coordinate transformations. Notice that the required counter term is not generated by the standard Lagrangian of the Salam-Weinberg model coupled minimally to the gravitational field.
The finite coupling constant of this non-minimal interaction term is a free parameter contributing to the $D$ term of the $Z$-boson at tree level. 

By considering matrix elements of the EMT operator for sharply localized states we interpret the obtained densities as spatial distributions of various quantities. 
In exact analogy to the case of the Higgs boson \cite{Beissner:2025nmg} we found that the $Z$-boson, as probed by the gravitational interaction, has non-zero extension. 
Our calculations resulted in the value of $(9.5\times 10^{-9} +8.2\times10^{-12} \, i) \ {\rm fm}^2$ for the mean-square energy radius of the $Z$-boson, which is about six times smaller than the corresponding quantity for the Higgs boson. The imaginary part of this quantity results from the considered system being unstable.

\acknowledgements

This work was supported in part 
by the MKW NRW under the funding code NW21-024-A, by the Georgian
Shota Rustaveli National 
Science Foundation (Grant No. FR-23-856), 
by the European Research Council (ERC) under the European Union's
Horizon 2020 research and innovation programme (grant agreement
No. 885150)
and by the EU Horizon 2020 research and
innovation programme (STRONG-2020, grant agreement No. 824093).

\appendix
\section{Definition of loop integrals}
\label{AppA}

One-loop integrals are defined as follows:
\begin{eqnarray}
 \text{A}_0(m^2)&=&\frac{(2\pi \mu)^{4-D}}{i\pi^2} \int\frac{d^D k}{k^2-m^2+i \epsilon}\,,\nonumber\\
 \text{B}_0(p^2,m_1^2,m_2^2)&=&\frac{(2\pi \mu)^{4-D}}{i\pi^2} \int\frac{d^Dk}{[k^2-m_1^2+i \epsilon] [(p+k)^2-m_2^2+i \epsilon]}\,, 
 \nonumber\\
C_0(p_1^2,p_2^2,p^2_{12},m_1^2,m_2^2,m_3^2)&=&\frac{(2\pi \mu)^{4-n}}{i\pi^2} \int\frac{d^nk}{[k^2-m_1^2+i \epsilon] [(p_1+k)^2-m_2^2+i \epsilon] [(p_1+p_2+k)^2-m_ 3^2+i \epsilon]}\,,
\label{ints}
\end{eqnarray}
where $p_{12}=p_1+p_2$ and $D$ is the spacetime dimension and $\mu$ the scale parameter.
For tensor loop integrals, we apply the reduction formulae of Ref.~\cite{Denner:2005nn},  
while for the expansion of the scalar integrals in Eq.~(\ref{ints}) in terms of kinematical invariants we use Ref.~\cite{Devaraj:1997es}.

\section{Contributions to the pole position and the residue of the
  Higgs propagator}
\label{AppB}

One-loop contributions to the pole position and the residue of the $Z$-boson propagator are given by
\begin{eqnarray}
\delta z_1 & = & \frac{e^2 A_0\left(m_n^2\right) \left((D-2) M_H^2 \left(4 M_Z^2 Q_n s_n
   \left(M_Z^2-M_W^2\right)+8 Q_n^2 \left(M_W^2-M_Z^2\right){}^2+M_Z^4\right)-4
   (D-1) m_n^2 M_Z^4\right)}{32 \pi ^2 (D-1) M_H^2 M_W^2
   \left(M_W^2-M_Z^2\right)} \nonumber\\
   &-& \frac{e^2 B_0\left(M_Z^2,m_n^2,m_n^2\right) }{64 \pi ^2 (D-1) M_W^2
   \left(M_W^2-M_Z^2\right)}
   \left((D-2) M_Z^2 \left(4 M_Z^2 Q_n s_n \left(M_Z^2-M_W^2\right)+8 Q_n^2
   \left(M_W^2-M_Z^2\right){}^2+M_Z^4\right) \right. \nonumber\\
   &-& \left. 2 m_n^2 \left((D-3) M_Z^4+8 M_Z^2
   Q_n s_n \left(M_W^2-M_Z^2\right)-16 Q_n^2
   \left(M_W^2-M_Z^2\right){}^2\right)\right) \nonumber\\
   &-& \frac{e^2 M_Z^2 \left(4 (D-1) M_Z^4-4 M_H^2 M_Z^2+M_H^4\right)
   B_0\left(M_Z^2,M_Z^2,M_H^2\right)}{64 \pi ^2 (D-1) M_W^2
   \left(M_W^2-M_Z^2\right)} \nonumber\\
   &+& \frac{e^2 \left(4 M_W^2-M_Z^2\right) \left(4 (2
   D-3) M_W^2 M_Z^2+4 (D-1) M_W^4+M_Z^4\right)
   B_0\left(M_Z^2,M_W^2,M_W^2\right)}{64 \pi ^2 (D-1) M_W^2
   \left(M_W^2-M_Z^2\right)} \nonumber\\
   &+& \frac{e^2 A_0\left(M_W^2\right) \left(-8
   \left(D^2-3 D+2\right) M_H^2 M_W^4+4 M_W^2 \left(2 (D-2) M_H^2 M_Z^2+(D-1)^2
   M_Z^4\right)+2 M_H^2 M_Z^4\right)}{64 \pi ^2 (D-1) M_H^2 M_W^2
   \left(M_W^2-M_Z^2\right)} \nonumber\\
   &+& \frac{e^2 M_Z^2 A_0\left(M_H^2\right) \left(2
   (D-1) M_Z^2+M_H^2\right)}{64 \pi ^2 (D-1) M_W^2
   \left(M_W^2-M_Z^2\right)}+\frac{e^2 A_0\left(M_Z^2\right) \left(2 (D-1)^2
   M_Z^6+2 M_H^2 M_Z^4-M_H^4 M_Z^2\right)}{64 \pi ^2 (D-1) M_H^2 M_W^2
   \left(M_W^2-M_Z^2\right)} \,,
   \label{deltaz1}
   \end{eqnarray}
   and
   \begin{eqnarray}
   \delta Z_1 &=& -\frac{e^2 B_0\left(M_Z^2,m_n^2,m_n^2\right)}{64 \pi ^2 M_W^2
   \left(M_W^2-M_Z^2\right)} \left(\frac{\left(2
   m_n^2+M_Z^2\right) \left(4 M_Z^2 Q_n s_n \left(M_Z^2-M_W^2\right)+8 Q_n^2
   \left(M_W^2-M_Z^2\right){}^2+M_Z^4\right)}{(D-1) M_Z^2} \right. \nonumber\\
   &-& \left. \frac{2 m_n^2
   \left(-2 m_n^2 M_Z^2+4 M_Z^2 Q_n s_n \left(M_Z^2-M_W^2\right)+8 Q_n^2
   \left(M_W^2-M_Z^2\right){}^2+M_Z^4\right)}{M_Z^2-4 m_n^2}-4 M_Z^2 Q_n s_n
   \left(M_Z^2-M_W^2\right) \right. \nonumber\\
   &-& \left. 8 Q_n^2
   \left(M_W^2-M_Z^2\right){}^2-M_Z^4\right)\nonumber\\
   &-& \frac{e^2 A_0\left(m_n^2\right) }{64 \pi ^2 M_W^2 \left(M_W^2-M_Z^2\right)} 
   \left(\frac{2
   \left(-2 m_n^2 M_Z^2+4 M_Z^2 Q_n s_n \left(M_Z^2-M_W^2\right)+8 Q_n^2
   \left(M_W^2-M_Z^2\right){}^2+M_Z^4\right)}{M_Z^2-4 m_n^2} \right. \nonumber\\ 
   &-& \left. \frac{2 \left(4
   M_Z^2 Q_n s_n \left(M_Z^2-M_W^2\right)+8 Q_n^2
   \left(M_W^2-M_Z^2\right){}^2+M_Z^4\right)}{(D-1) M_Z^2}\right) \nonumber\\
   &-& \frac{e^2 \left(M_Z^2-2 m_n^2\right) }{64 \pi ^2 (D-1) M_W^2 M_Z^2
   \left(M_W^2-M_Z^2\right) \left(M_Z^2-4 m_n^2\right)} 
   \left((D-2) M_Z^2 \left(4 M_Z^2 Q_n s_n \left(M_Z^2-M_W^2\right) \right. \right. \nonumber\\
   &+& \left. \left. 8 Q_n^2
   \left(M_W^2-M_Z^2\right){}^2+M_Z^4\right)-2 m_n^2 \left((D-3) M_Z^4 
   + 8 M_Z^2
   Q_n s_n \left(M_W^2-M_Z^2\right)-16 Q_n^2
   \left(M_W^2-M_Z^2\right){}^2\right)\right)\nonumber\\
   &+& \frac{ e^2}{64 (D-1)
   \pi ^2 M_W^2 M_Z^2 \left(M_W^2-M_Z^2\right) \left(4
   M_Z^2-M_H^2\right)}  \left(8 (D-1) \left(4 M_Z^2-M_H^2\right) M_W^6 \right. \nonumber\\
   &+& \left. 4 (3 D-5) M_Z^2 \left(4
   M_Z^2-M_H^2\right) M_W^4-2 (4 D-7) M_Z^4 \left(4 M_Z^2-M_H^2\right)
   M_W^2-M_Z^4 \left(M_H^4-5 M_Z^2 M_H^2+4 D M_Z^4\right)\right) \nonumber\\
   &+& \frac{\left(-2 M_H^6+11 M_Z^2 M_H^4-4 (D+2) M_Z^4
   M_H^2+8 (D-1) M_Z^6\right) A_0\left(M_H^2\right) e^2}{64 (D-1) \pi ^2 M_H^2
   M_W^2 \left(M_W^2-M_Z^2\right) \left(4
   M_Z^2-M_H^2\right)} \nonumber\\
   &-& \frac{\left(\left(4 D-2 \left(\alpha _W+1\right)\right)
   M_W^4+2 M_Z^2 \left(6 D+\alpha _W-10\right) M_W^2+M_Z^4\right)
   A_0\left(M_W^2\right) e^2}{32 (D-1) \pi ^2 M_W^2 M_Z^2
   \left(M_W^2-M_Z^2\right)}  \nonumber\\
   &+&\frac{\left(2 M_H^4-9 M_Z^2 M_H^2+4 D M_Z^4\right)
   A_0\left(M_Z^2\right) e^2}{64 (D-1) \pi ^2 M_W^2 \left(M_W^2-M_Z^2\right)
   \left(4 M_Z^2-M_H^2\right)}-\frac{\left(M_Z^2 \left(2 D+\alpha
   _W-4\right)-M_W^2 \left(\alpha _W-1\right)\right) A_0\left(M_W^2 \alpha
   _W\right) e^2}{16 (D-1) \pi ^2 M_Z^2 \left(M_Z^2-M_W^2\right)}\nonumber\\
   &+& \frac{\left(8
   (D-1) M_W^6-4 (3 D-5) M_Z^2 M_W^4+2 (8 D-15) M_Z^4 M_W^2+3 M_Z^6\right)
   B_0\left(M_Z^2,M_W^2,M_W^2\right) e^2}{64 (D-1) \pi ^2 M_W^2 M_Z^2
   \left(M_W^2-M_Z^2\right)}\nonumber\\
   &+& \frac{\left(\left(\alpha _W-1\right){}^2 M_W^4+2
   M_Z^2 \left(2 D-\alpha _W-3\right) M_W^2+M_Z^4\right)
   B_0\left(M_Z^2,M_W^2,M_W^2 \alpha _W\right) e^2}{16 (D-1) \pi ^2 M_W^2
   M_Z^2}\nonumber\\
   &-& \frac{\left(-2 M_H^6+13 M_Z^2 M_H^4-4 (D+4) M_Z^4 M_H^2+12 (D-1)
   M_Z^6\right) B_0\left(M_Z^2,M_Z^2,M_H^2\right) e^2}{64 (D-1) \pi ^2 M_W^2
   \left(M_W^2-M_Z^2\right) \left(4 M_Z^2-M_H^2\right)} \nonumber\\
   &+& \frac{M_Z^2
   \left(M_Z^2-4 M_W^2 \alpha _W\right) B_0\left(M_Z^2,M_W^2 \alpha _W,M_W^2
   \alpha _W\right) e^2}{32 (D-1) \pi ^2 M_W^2 \left(M_W^2-M_Z^2\right)}
   \,.
\label{ZSE}
\end{eqnarray}

\section{Gravitational form factors at $q^2=0$}
\label{GFFs0}

The values of the form factors at zero momentum-transfer have
the form:
\begin{eqnarray}
A_0(0) & = & 1\,,\nonumber\\
A_1(0) & = & \frac{e^2 m_n^4 M_V^2 B_0\left(M_Z^2,m_n^2,m_n^2\right) \left(4 M_Z^2 Q_n s_n
   \left(M_Z^2-M_W^2\right)+8 Q_n^2 \left(M_W^2-M_Z^2\right){}^2+M_Z^4\right)}{12
   \pi ^2 M_W^2 M_Z^4 \left(M_W^2-M_Z^2\right) \left(M_Z^2-4
   m_n^2\right)} \nonumber\\
   &-& \frac{e^2 m_n^2 M_V^2 A_0\left(m_n^2\right) \left(4 M_Z^2 Q_n
   s_n \left(M_Z^2-M_W^2\right)+8 Q_n^2
   \left(M_W^2-M_Z^2\right){}^2+M_Z^4\right)}{12 \pi ^2 M_W^2 M_Z^4
   \left(M_W^2-M_Z^2\right) \left(M_Z^2-4 m_n^2\right)} \nonumber\\
   &-& \frac{e^2 M_V^2 \left(2
   m_n^2 M_Z^2-12 m_n^4+M_Z^4\right) \left(4 M_Z^2 Q_n s_n
   \left(M_Z^2-M_W^2\right)+8 Q_n^2
   \left(M_W^2-M_Z^2\right){}^2+M_Z^4\right)}{144 \pi ^2 M_W^2 M_Z^4
   \left(M_W^2-M_Z^2\right) \left(M_Z^2-4 m_n^2\right)}\nonumber\\
&-& \frac{e^2 M_V^2 \left(6 M_H^4 M_Z^2-9 M_H^2 M_Z^4-M_H^6+2 M_Z^6\right)
   B_0\left(M_Z^2,M_Z^2,M_H^2\right)}{24 \pi ^2 M_W^2 M_Z^2 \left(4
   M_Z^2-M_H^2\right) \left(M_W^2-M_Z^2\right)}\nonumber\\
   &+& \frac{e^2 M_V^2 M_W^2 \left(-4
   M_W^2 M_Z^2+12 M_W^4+M_Z^4\right) B_0\left(M_Z^2,M_W^2,M_W^2\right)}{12 \pi ^2
   M_Z^4 \left(-5 M_W^2 M_Z^2+4 M_W^4+M_Z^4\right)}
   -\frac{e^2 M_V^2
   A_0\left(M_H^2\right) \left(-5 M_H^2 M_Z^2+M_H^4+5 M_Z^4\right)}{24 \pi ^2
   M_W^2 M_Z^2 \left(4 M_Z^2-M_H^2\right) \left(M_W^2-M_Z^2\right)} \nonumber\\
   &+&\frac{e^2
   M_V^2 A_0\left(M_Z^2\right) \left(-4 M_H^2 M_Z^2+M_H^4+2 M_Z^4\right)}{24 \pi
   ^2 M_W^2 M_Z^2 \left(4 M_Z^2-M_H^2\right) \left(M_W^2-M_Z^2\right)} -\frac{e^2
   M_V^2 A_0\left(M_W^2\right) \left(-4 M_W^2 M_Z^2+12 M_W^4+M_Z^4\right)}{12 \pi
   ^2 M_Z^4 \left(-5 M_W^2 M_Z^2+4 M_W^4+M_Z^4\right)} \nonumber\\
   &+& \frac{e^2 M_V^2 }{144 \pi ^2
   M_W^2 M_Z^4 \left(4 M_Z^2-M_H^2\right) \left(M_W^2-M_Z^2\right) \left(4
   M_W^2-M_Z^2\right)}  \left(144
   M_W^8 \left(4 M_Z^2-M_H^2\right)-72 M_W^6 \left(4 M_Z^4-M_H^2 M_Z^2\right) \right. \nonumber\\
   &+& \left. 8
   M_W^4 \left(4 M_Z^6-M_H^2 M_Z^4\right)+6 M_W^2 \left(5 M_H^2 M_Z^6-2 M_H^4
   M_Z^4+4 M_Z^8\right)-7 M_H^2 M_Z^8+3 M_H^4 M_Z^6-8 M_Z^{10}\right) 
   \,,
   \label{A10}
\end{eqnarray}   
\begin{eqnarray}
J(0) & = & 1\,,\nonumber\\
D_0(0) & = & 1+ \frac{e^2 m_n^2 B_0\left(M_Z^2,m_n^2,m_n^2\right) \left(8 M_Z^2 Q_n s_n
   \left(M_Z^2-M_W^2\right)+16 Q_n^2
   \left(M_W^2-M_Z^2\right){}^2+M_Z^4\right)}{48 \pi ^2 M_W^2
   \left(M_W^2-M_Z^2\right) \left(M_Z^2-4 m_n^2\right)} 
   \nonumber\\
   &-& \frac{e^2
   A_0\left(m_n^2\right) \left(8 M_H^2 M_Z^2 Q_n s_n \left(M_Z^2-M_W^2\right)+16
   M_H^2 Q_n^2 \left(M_W^2-M_Z^2\right){}^2+M_Z^4 \left(M_H^2+2 M_Z^2\right)-8
   m_n^2 M_Z^4\right)}{48 \pi ^2 M_H^2 M_W^2 \left(M_W^2-M_Z^2\right)
   \left(M_Z^2-4 m_n^2\right)} \nonumber\\
   &+& \frac{e^2 }{48 \pi ^2 M_H^2 M_W^2 \left(M_W^2-M_Z^2\right)
   \left(M_Z^2-4 m_n^2\right)} \left(m_n^2 \left(8 M_H^2 M_Z^2 Q_n s_n
   \left(M_Z^2-M_W^2\right)+16 M_H^2 Q_n^2 \left(M_W^2-M_Z^2\right){}^2 \right.  \right.\nonumber\\
   &+& \left. \left. 3 M_H^2
   M_Z^4+2 M_Z^6\right)-M_H^2 M_Z^2 \left(4 M_Z^2 Q_n s_n
   \left(M_Z^2-M_W^2\right)+8 Q_n^2 \left(M_W^2-M_Z^2\right){}^2+M_Z^4\right)-8
   m_n^4 M_Z^4\right)\nonumber\\
   &+& \frac{e^2 M_Z^4 \left(M_H^2-2 M_Z^2\right){}^2
   B_0\left(M_Z^2,M_Z^2,M_H^2\right)}{48 \pi ^2 M_H^2 M_W^2 \left(4
   M_Z^2-M_H^2\right) \left(M_W^2-M_Z^2\right)}+\frac{e^2 \left(-4 M_W^2 M_Z^2+12
   M_W^4-M_Z^4\right) B_0\left(M_Z^2,M_W^2,M_W^2\right)}{48 \pi ^2 M_W^2
   \left(M_W^2-M_Z^2\right)} \nonumber\\
   &+& \frac{e^2 M_Z^4 A_0\left(M_H^2\right) \left(M_H^2-6
   M_Z^2\right)}{48 \pi ^2 M_H^2 M_W^2 \left(4 M_Z^2-M_H^2\right)
   \left(M_W^2-M_Z^2\right)}+\frac{e^2 A_0\left(M_W^2\right) \left(-6 M_H^2
   M_W^2+2 M_H^2 M_Z^2+3 M_Z^4\right)}{24 \pi ^2 M_H^2 M_W^2
   \left(M_W^2-M_Z^2\right)} \nonumber\\
   &+&  \frac{e^2  \left(12 M_W^4 \left(4 M_H^2
   M_Z^2-M_H^4\right)-2 M_W^2 \left(-25 M_H^2 M_Z^4+6 M_H^4 M_Z^2+4
   M_Z^6\right)-M_Z^4 \left(-5 M_H^2 M_Z^2+M_H^4+8 M_Z^4\right)\right) }{48 \pi ^2
   M_H^2 M_W^2 \left(4 M_Z^2-M_H^2\right) \left(M_W^2-M_Z^2\right)} \nonumber\\
   &+& \frac{e^2 A_0\left(M_Z^2\right) \left(8 M_Z^6-M_H^2
   M_Z^4\right)}{48 \pi ^2 M_H^2 M_W^2 \left(4 M_Z^2-M_H^2\right)
   \left(M_W^2-M_Z^2\right)} 
 \,, 
    \label{D00}
\end{eqnarray}   
\begin{eqnarray}
   D_1(0) & = & \frac{e^2 M_V^2 \left(-5 m_n^2 M_Z^2+5 m_n^4+M_Z^4\right)
   B_0\left(M_Z^2,m_n^2,m_n^2\right) \left(4 M_Z^2 Q_n s_n
   \left(M_Z^2-M_W^2\right)+8 Q_n^2 \left(M_W^2-M_Z^2\right){}^2+M_Z^4\right)}{60
   \pi ^2 M_W^2 M_Z^4 \left(M_W^2-M_Z^2\right) \left(M_Z^2-4
   m_n^2\right)} \nonumber\\
   &-& \frac{e^2 M_V^2 A_0\left(m_n^2\right) \left(-5 m_n^2 M_Z^2+5
   m_n^4+M_Z^4\right) \left(4 M_Z^2 Q_n s_n \left(M_Z^2-M_W^2\right)+8 Q_n^2
   \left(M_W^2-M_Z^2\right){}^2+M_Z^4\right)}{60 \pi ^2 m_n^2 M_W^2 M_Z^4
   \left(M_W^2-M_Z^2\right) \left(M_Z^2-4 m_n^2\right)} \nonumber\\
   &+& \frac{e^2 M_V^2 \left(-70
   m_n^2 M_Z^2+60 m_n^4+13 M_Z^4\right) \left(4 M_Z^2 Q_n s_n
   \left(M_Z^2-M_W^2\right)+8 Q_n^2
   \left(M_W^2-M_Z^2\right){}^2+M_Z^4\right)}{720 \pi ^2 M_W^2 M_Z^4
   \left(M_W^2-M_Z^2\right) \left(M_Z^2-4 m_n^2\right)}  \nonumber\\
   &+& \frac{e^2 M_V^2}{720 \pi ^2 M_H^2 M_W^2 M_Z^4
   \left(M_W^2-M_Z^2\right) \left(4 M_W^2-M_Z^2\right) \left(4
   M_Z^2-M_H^2\right)}  \left(720 \left(4 M_H^2 M_Z^2-M_H^4\right) M_W^8 \right. \nonumber\\
   &-& \left. 1800 \left(4 M_H^2
   M_Z^4-M_H^4 M_Z^2\right) M_W^6-104 \left(4 M_H^2 M_Z^6-M_H^4 M_Z^4\right)
   M_W^4+6 M_Z^4 \left(-10 M_H^6-3 M_Z^2 M_H^4 \right. \right. \nonumber\\
   &+& \left. \left. 172 M_Z^4 M_H^2+32 M_Z^6\right)
   M_W^2-M_Z^6 \left(-15 M_H^6+23 M_Z^2 M_H^4+148 M_Z^4 M_H^2+48
   M_Z^6\right)\right) \nonumber\\
   &-& \frac{e^2 \left(5 M_H^6-25 M_Z^2 M_H^4+15 M_Z^4 M_H^2+16
   M_Z^6\right) A_0\left(M_H^2\right) M_V^2}{120 \pi ^2 M_H^2 M_W^2 M_Z^2
   \left(M_W^2-M_Z^2\right) \left(4 M_Z^2-M_H^2\right)} \nonumber\\
   &-& \frac{e^2 \left(60
   M_W^8-140 M_Z^2 M_W^6-3 M_Z^4 M_W^4+26 M_Z^6 M_W^2-4 M_Z^8\right)
   A_0\left(M_W^2\right) M_V^2}{60 \pi ^2 M_W^4 M_Z^4 \left(4 M_W^4-5 M_Z^2
   M_W^2+M_Z^4\right)} \nonumber\\
   &+& \frac{e^2 \left(5 M_H^6-20 M_Z^2 M_H^4-4 M_Z^4 M_H^2+8
   M_Z^6\right) A_0\left(M_Z^2\right) M_V^2}{120 \pi ^2 M_H^2 M_W^2 M_Z^2
   \left(M_W^2-M_Z^2\right) \left(4 M_Z^2-M_H^2\right)} \nonumber\\
   &+& \frac{e^2 \left(60
   M_W^8-140 M_Z^2 M_W^6-3 M_Z^4 M_W^4+26 M_Z^6 M_W^2-4 M_Z^8\right)
   B_0\left(M_Z^2,M_W^2,M_W^2\right) M_V^2}{60 \pi ^2 M_W^2 M_Z^4 \left(4 M_W^4-5
   M_Z^2 M_W^2+M_Z^4\right)} \nonumber\\
   &-& \frac{e^2 \left(-5 M_H^8+30 M_Z^2 M_H^6-35 M_Z^4
   M_H^4-20 M_Z^6 M_H^2+8 M_Z^8\right) B_0\left(M_Z^2,M_Z^2,M_H^2\right)
   M_V^2}{120 \pi ^2 M_H^2 M_W^2 M_Z^2 \left(M_W^2-M_Z^2\right) \left(4
   M_Z^2-M_H^2\right)} \,, 
\label{D10}
\end{eqnarray}   
\begin{eqnarray}   
  E(0) & = & 1 + \frac{e^2 m_n^2 B_0\left(M_Z^2,m_n^2,m_n^2\right) }{96 \pi ^2 M_W^2 M_Z^2 \left(M_W^2-M_Z^2\right)
   \left(M_Z^2-4 m_n^2\right)} \left(2 m_n^2 \left(16 M_Z^2
   Q_n s_n \left(M_Z^2-M_W^2\right)+32 Q_n^2
   \left(M_W^2-M_Z^2\right){}^2+M_Z^4\right) \right. \nonumber\\
   &+& \left. 4 M_Z^4 Q_n s_n
   \left(M_Z^2-M_W^2\right)+8 Q_n^2 \left(M_Z^3-M_W^2
   M_Z\right){}^2+M_Z^6\right) \nonumber\\
   &-& \frac{e^2 A_0\left(m_n^2\right) }{96 \pi ^2 M_W^2 M_Z^2 \left(M_W^2-M_Z^2\right)
   \left(M_Z^2-4 m_n^2\right)} 
    \left(2 m_n^2
   \left(16 M_Z^2 Q_n s_n \left(M_Z^2-M_W^2\right)+32 Q_n^2
   \left(M_W^2-M_Z^2\right){}^2+M_Z^4\right) \right. \nonumber\\
   &+& \left. 4 M_Z^4 Q_n s_n
   \left(M_Z^2-M_W^2\right)+8 Q_n^2 \left(M_Z^3-M_W^2
   M_Z\right){}^2+M_Z^6\right)\nonumber\\
   &+& \frac{e^2 }{576 \pi ^2 M_W^2 M_Z^2
   \left(M_W^2-M_Z^2\right) \left(M_Z^2-4 m_n^2\right)} 
   \left(12 m_n^4 \left(16 M_Z^2 Q_n s_n
   \left(M_Z^2-M_W^2\right)+32 Q_n^2 \left(M_W^2-M_Z^2\right){}^2+M_Z^4\right) \right. \nonumber\\
   &-& \left. 
   8 m_n^2 \left(M_Z^4 Q_n s_n \left(M_Z^2-M_W^2\right)+2 Q_n^2 \left(M_Z^3-M_W^2
   M_Z\right){}^2-2 M_Z^6\right)-7 \left(4 M_Z^6 Q_n s_n
   \left(M_Z^2-M_W^2\right). \right. \right. \nonumber\\
   &+& \left. \left. 8 M_Z^4 Q_n^2
   \left(M_W^2-M_Z^2\right){}^2+M_Z^8\right)\right) \nonumber\\
   &-&  \frac{e^2 \left(13 M_H^4 M_Z^2-27 M_H^2 M_Z^4-2 M_H^6+16 M_Z^6\right)
   B_0\left(M_Z^2,M_Z^2,M_H^2\right)}{96 \pi ^2 M_W^2 \left(4 M_Z^2-M_H^2\right)
   \left(M_W^2-M_Z^2\right)} \nonumber\\
   &+& \frac{e^2 \left(-8 M_W^2 M_Z^2-12 M_W^4+5
   M_Z^4\right) B_0\left(M_Z^2,M_W^2,M_W^2\right)}{48 \pi ^2 M_Z^2
   \left(M_Z^2-M_W^2\right)}-\frac{e^2 A_0\left(M_H^2\right) \left(-11 M_H^2
   M_Z^2+2 M_H^4+18 M_Z^4\right)}{96 \pi ^2 M_W^2 \left(4 M_Z^2-M_H^2\right)
   \left(M_W^2-M_Z^2\right)} \nonumber\\
   &+& \frac{e^2 A_0\left(M_Z^2\right) \left(-9 M_H^2
   M_Z^2+2 M_H^4+16 M_Z^4\right)}{96 \pi ^2 M_W^2 \left(4 M_Z^2-M_H^2\right)
   \left(M_W^2-M_Z^2\right)}-\frac{e^2 A_0\left(M_W^2\right) \left(8 M_W^2
   M_Z^2+12 M_W^4-5 M_Z^4\right)}{48 \pi ^2 M_W^2 M_Z^2
   \left(M_W^2-M_Z^2\right)} \nonumber\\
   &+& \frac{e^2 }{288 \pi
   ^2 M_W^2 M_Z^2 \left(4 M_Z^2-M_H^2\right) \left(M_W^2-M_Z^2\right)}  \left(72 M_W^6 \left(4
   M_Z^2-M_H^2\right)+36 M_W^4 \left(4 M_Z^4-M_H^2 M_Z^2\right) \right. \nonumber\\
   &+& \left. 10 M_W^2 \left(4
   M_Z^6-M_H^2 M_Z^4\right)+11 M_H^2 M_Z^6-3 M_H^4 M_Z^4-32 M_Z^8\right) \,, 
      \label{E0}
\end{eqnarray}   
\begin{eqnarray}
   F_1(0) & = & -\frac{e^2 M_Z^2 \left(4 Q_n s_n \left(M_Z^2-M_W^2\right)+M_Z^2\right) \left(-2
   m_n^2 \left(B_0\left(M_Z^2,m_n^2,m_n^2\right)+1\right)+2
   A_0\left(m_n^2\right)+M_Z^2\right)}{64 \pi ^2 M_W^2
   \left(M_W^2-M_Z^2\right) \left(M_Z^2-4 m_n^2\right)} \,, 
\nonumber\\
   F_2(0) & = & 0
   \,, 
\nonumber\\
   F_3(0) & = & \frac{e^2 M_Z^2 \left(4 Q_n s_n \left(M_Z^2-M_W^2\right)+M_Z^2\right)
   \left(-2 m_n^2 \left(B_0\left(M_Z^2,m_n^2,m_n^2\right)+1\right)+2
   A_0\left(m_n^2\right)+M_Z^2\right)}{128 \pi ^2 M_W^2
   \left(M_W^2-M_Z^2\right) \left(M_Z^2-4 m_n^2\right)} \,, 
\nonumber\\
   F_4(0) & = & \frac{e^2 M_Z^2 \left(4 Q_n s_n \left(M_Z^2-M_W^2\right)+M_Z^2\right) \left(-2
   m_n^2 \left(B_0\left(M_Z^2,m_n^2,m_n^2\right)+1\right)+2
   A_0\left(m_n^2\right)+M_Z^2\right)}{128 \pi ^2 M_W^2
   \left(M_W^2-M_Z^2\right) \left(M_Z^2-4 m_n^2\right)} \,, 
\nonumber\\
   F_5(0) & = & 0
    \,.
\label{Fs0}
\end{eqnarray}

\end{document}